\documentclass{style/llncs}
\usepackage{url}
\usepackage{amsmath}
\usepackage{amssymb}
\usepackage{graphicx}
\usepackage{tabularx}
\usepackage[numbers]{natbib}
\usepackage{floatrow}
\usepackage{listings}
\makeatletter
\renewcommand\bibsection%
{
	\section*{\refname
		\@mkboth{\MakeUppercase{\refname}}{\MakeUppercase{\refname}}}
}
\makeatother

\begin{document}
	
	\title{Agile Behaviour Design: A Design Approach for Structuring Game Characters and Interactions}
	

	\author{ Swen E. Gaudl}
	\institute{Falmouth University, MetaMakers Institute \email{swen.gaudl@gmail.com} }

	\maketitle
	\begin{abstract}
		In this paper, a novel design methodology---\textsc{Agile Behaviour Design}---is presented that accommodates the requirements for developing complex game agents suitable for industrial environments. An essential part of the design approach is to support independent work of both designers and programmers by reducing bottleneck situations. The approach fosters the creation of more loose and fluid interactions between design and implementation, leaving more freedom for creative expression. 
	\end{abstract}
	
	
	\section{Introduction}
	In game development and similarly in other dynamic software projects, \textsc{scrum} is the dominating approach \cite{rubin2012essential,keith2010agile} for developing products in a managed way. The \textsc{scrum} process model is based on the agile philosophy, supporting alterations to the product late in the project; a situation often required in game development. ``\textsc{scrum} for games'' \cite{keith2010agile}, an approach specifically adjusted for games, discusses four phases which partition the development, i.e. concept, pre-production, production, post-production. The phases have defined milestone points that are generated at the start of each project. Even for non-industrial game development or game AI design, process models are useful. They provide a common framework that can support the integrating different parties into a shared project that converges on a common goal. However, ``\textsc{scrum} for games'' is described only on a high abstraction level not aiding the design of specific game components such as the game AI system.
	In contrast to \textsc{scrum} which focuses only on a high-level approach, game character development requires the integration of lower level design and approaches as well. Thus, for story generation which often includes the low level design of character AI driving the story, an approach is needed that integrates both high and low level design. This paper introduces \textsc{Agile Behaviour Design (A-BeD)} an agile design methodology that embeds both levels into its approach including the need for a better author integration. The approach also provides another solution to the authoring problem for interactive storytelling \cite{spierling2009authoring}.
	%
	%
	Traditional game development phases impact and direct the creative process and freedom of designers and influence the system design. The more mature the system or game becomes, the more restrictive are deviations from the initial design. Thus, features need to be known early and should not emerge late in the design. This affects character and interaction design massively as alterations to them emerge during interacting with the system and user testing. In those instances, \textsc{A-BeD} can support the process and the development of behaviour-based agent AI responsible for creating dynamic and "living" stories.
	
	\section{Background \& Related Work} \label{sec:bg}
	\textsc{A-BeD} is based on the \textsc{Behavior-Oriented Design (bod)} methodology\cite{Bryson-PHD}, a top-down analysis of a desired behaviour combined with a bottom-up generation of plans and a behaviour library\footnote{Behaviour plans are designed to be human readable/amendable. The behaviour library is compiled game engine or agent framework specific program code.}. The \textsc{bod} top-down analysis starts with the definition of a high-level task the agent wants to achieve, an undertaking for generating a single agent in a well-defined environment by an expert. Next, the plans are built bottom-up by implementing primitives and enriching the behaviour plan. Primitives are clustered into behaviours according to their usage of shared memory/state objects. New goals and sub-trees are added until the agent is capable of performing the initially defined task. An existing systems analysis, comparing three \textsc{Intelligent Virtual Agent} (IVA) approaches---\textsc{bod}\cite{Bryson-PHD}, \textsc{ABL}\cite{Mateas2002} \& \textsc{FaTiMa} \cite{dias2014fatima} , found that the decomposition and plan creation to be a challenging process \cite{grow2014methodology}. The process requires authors to able to express themselves within a specific tool while at the same time maintaining their creative design goal under the system's restrictions. 
	Supporting this process is crucial for reducing the burden on novice or less technical users. Similar observations were made during an undergraduate AI course, taught at the University of Bath, where student as part of their coursework created \textsc{iva}s using \textsc{bod}. Novices tended to generate either flat shallow plans or deep narrow plans, restricting the resulting agent immensely. These observations also apply to other approaches such as \textsc{BehaviorTree (bt)} \cite{Champandard2013} as they are part of authorial burdens \cite{spierling2009authoring}.  
	When using \textsc{bod}/creating a \textbf{[BT]}, iterating over the plan\textbf{[tree]} and creating behaviour primitives\textbf{[nodes]} results in a tight coupling of programmer and designer as the entire behaviour library and plan structure is in flux. This is undesirable as it locks both parties into very restrictive patterns of interaction.
	%
	Additionally, increased agent sizes or story depth such as the narrative in Fa\c{c}ade \cite{Mateas2002a} lead to growing complexity of the underlying structure as well which requires a lot of careful design. Thus, selecting the right system architectures and approach is important. A system based on finite-state machines increases its complexity exponentially, even in the average case \textit{(normal design)}, which renders any system at a certain stage unusable. Frameworks for modelling behaviour such as \textsc{bt} and \textsc{posh} \cite{Gaudl2013,Gaudl-PHD} have an exponential complexity growth only in the worst case \textit{(bad design)}. Planning systems such as \textsc{goap} \citep{orkin2005agent} require authors able program expert systems as the plan they code is highly complex. However, this reduces the interdependence of nodes and the amount of manual work, e.g. checking all transitions. \textsc{posh} integrates a lightweight planner allowing local design by modifying existing sub-trees and hierarchically nesting them within its modular structure.

	\section{A Directed Model for Behaviour Design}
	To advance \textsc{bod} into \textsc{A-BeD}, elements of the \textsc{scrum} process were integrated to guide the agent design and the new process was designed to converge more against a final product.
	\textsc{Scrum} is an agile software development process integrating iterative development and testing while maintaining as much as possible the time predictability from other development processes such as the Waterfall model. It partitions the project into smaller \emph{Sprints}, each taking a specified time and dealing with a defined set of features/tasks. At the end of each \emph{Sprint}, the entire system should be able to execute the features developed during the Sprint, including those that have been newly integrated. Features are collected on a feature board which presents them in ordered lists (product backlog) of completed, in-progress and to-be-implemented elements. "Scrum for Games" \cite{keith2010agile} starts with an initial full system specification and continuous stable versions of the product while incrementally adding features from a feature board. The important part,the feature board, is created and laid out to schedule the work and progress of all features. This contrasts conventional \textsc{scrum} where work is scheduled into tasks that can contain partial or parts of multiple features. After all features for the final product and production phases have been laid out, the implementation starts. The starting point of \textsc{bod} is initially a minimal plan containing a small set of action primitives,as shown in \cite{Partington05,Bryson-PHD}. 
	
	\subsection{Agile Behaviour Design}
	
	The first step is to decompose a given scenario \textit{(1)} into a full set of behaviour primitives( i.e., actions and senses) and state variables. In the case of a stealth game, those could include moving to a location, sensing if a player is close or opening/closing a door. 
	After that, the designer is building a full behaviour plan \textit{(2)} for the agent that suffices the scenario specification. This step is more time-consuming than the incremental build up using \textsc{bod} and cognitively more challenging. It should be done in as few sessions as possible by building an entire behaviour plan bottom up using the previously specified primitives. This part of the development is a pure design task without the need for programmer involvement. Using a planning system this would involve noting all pre-\&post-conditions and states, e.g. if the player is visible and the NPC is too far away it moves closer and triggers a dialogue.
	Next, the initial design plan is evaluated \textit{(3)} together with a programmer. Based on the feedback, the design is modified; primitives are added, adjusted and renamed.
	Behaviour stubs are generated \textit{(4a)} in an \textsc{Object-Oriented Design (OOD)} fashion. All specified primitive are stubbed and clustered into them according to memory/state usage. This stage is a pure programmer task as it involves creating empty methods such as \textit{moveTo(Location), sense(Player), open(Object)} which are referenced in the design.
	New primitives should contain a default return state \textit{(4b)}. At this point, the fallback action should be called if other plan elements fail. It is the only action which needs an implementation in the beginning. This primitive allows the plan to be executable and represents an \textit{idle} state of the agent.
	When designing a behaviour plan, its sub-plans (sub-trees in \textsc{bt}) are triggered upon meeting one or multiple conditions. Thus, when conditions are not fulfilled, the trigger does not release the related sub-tree. Following the previous example, an agent might not be able to \textit{sense} the player if a door is closed. Thus, the sub-tree dealing with the player will not be triggered. Using this mechanism, it is possible to deactivate parts of the plan similar to the bitmasks used by \citeauthor{isla_gdc_2005} \cite{isla_gdc_2005}. To achieve this, the designer can integrate senses that unlock sub-trees if they are triggered. Once implemented, those senses can activate the sub-plan. Interacting with the player can be disabled by requiring the \textit{sense(Player)} to return \textit{true} and using a default /textit{false} return value to disable the sub-tree.
	
	After obtaining a first feature-complete plan, the work on the underlying behaviour primitives can be adjusted to work on individual features. Thus, the feature board can be ordered \textit{(5)} by clustering actions and senses under specific feature groups such as NPC movement, dialogue system, or combat.The alteration to the feature board can be done by grouping actions and senses according to their position in the hierarchical tree. This supports the identification of redundant or re-usable functionality by identifying similar usage of actions and senses within competences. 
	On the feature board, the relating features should be ordered so that sub-trees can be completed one at a time, thus, unlocking them for the agent. This clustering allows programmers to shift entire feature blocks up and down on the feature board without impacting other sub-trees.
	
	If the behaviour designer now decides to alter the plan, a large number of actions and senses are already stubbed within the hollow behaviour set. This given structure allows the designer to work independently \textit{(6b)} on the design while programmers can implement the stubs \textit{(6a)}. Following this approach requires fewer inclusions of new underlying primitives than following a simple incremental approach; it also distributes the work better between designer and programmer by initially close coordination in the first phase and a looser coupling later on.
	
	Ideally, the work is directed from bottom to top of plan following the idea of the \textsc{Subsumption} design \cite{brooks_robust_1986}. This will enable higher level drives after lower level ones have been implemented and tested. By approaching the iterative design \textit{(7)} this way, the complexity and expressiveness of the agent increases according to the designed priorities without impacting the robustness or completeness of the behaviour plan.
	
	\subsection{Agile Process Steps}
	\begin{enumerate}
		\item[1\hspace*{1.3ex}] Decompose scenario behaviour into primitives, states and goals (Design)
		\item[2\hspace*{1.3ex}] Design full behaviour plan that would suffice intended scenario (Design)
		\item[3\hspace*{1.3ex}] adjust/alter plan and primitive list (Design+Programming)
		\item[4a] Templating behaviour stubs (Programming)
		\item[4b] Design behaviour plan to have feature locks (Design)
		\item[5\hspace*{1.3ex}] Modify feature board \& sort according to sub-trees (Design+Programming)
		\item[6a] Implement stubs and alter primitives according to features (Programming)
		\item[6b] Test \& Develop behaviour plan based on given primitives/features (Design)
		\item[7\hspace*{1.3ex}] Loop to 3) until feature board is empty
	\end{enumerate}
	
	\noindent\textsc{Agile Behaviour Design} was used to develop agents for \textsc{starcraft} \cite{Gaudl-PHD} and in the development of the Android game \textsc{Stealthier POSH}\footnote{The game is available on the Android app store or using the following link: \url{https://play.google.com/store/apps/details?id=com.fairrats.POSH}} as a proof of concept. 
	Maintaining a prioritised feature set which relates to the plan sub-trees proved in those two case studies beneficial and provides transferable knowledge for commercial development environments as well. The feature board allows for better tracking of the development progress and more independent work of designers and programmers. 
	Additionally, it removes the burden of numerous changes to the behaviour library early in the project or restricting the designer from working purely on the plan without being able to test it. Unimplemented actions returning the default state allow for the parallel work on partial behaviours which decouples the programmer. 
	Using \textsc{A-BeD} with hierarchical planners such as \textsc{posh} enables working on smaller sections of an agent and concentrating for example on interaction with other agents while the dependencies between designer and programmer are reduced. Additionally, integrating the default trigger states, sub-trees unlock based on the progress of their underlying implementation. This cascaded unlocking of the tree and the resulting behaviour allows for a better version control of the behaviour library because it is more directed towards realising connected sub-trees. The combination of working on sub-trees and the feature board based on \textsc{scrum} directs the agent implementation to focus on connected pieces. The new approach should provide sufficient support for working on more complex systems or distributing work between different people such as movement and narration design for a given agent. 
	
	\section{Conclusion \& Future Work}
	This paper presents a novel, project-oriented methodology, extending the existing \textsc{behavior oriented design (bod)} \cite{Bryson-PHD}. One focus of the new methodology is to provide better separation of design and programming and to support the development of artificial agents in teams of multi-disciplinary authors. The two case studies and the feedback from the systems analysis \cite{grow2014methodology} create the basis for the newly introduced process steps of the methodology. The new process allows designers and programmers to distribute their work better while still following keeping the project progress in mind. \textsc{Agile Behaviour Design} reduces the dependencies of the different user groups such as authors, designers and programmers. To further aid the development and to focus on multi-platform development the arbitration architecture \textsc{posh-sharp} \cite{Gaudl-PHD} was designed to support specifically the agile design approach better.
	As a next step, further evaluations of the new methodology and approach are intended with novice and expert 
	users as well as a widened systematic analysis of development approaches to support cross-disciplinary design.
	%

	\bibliographystyle{style/splncsnat}
	\bibliography{biblio/sweneg,biblio/swen-ai-lit,biblio/swen-game-lit,biblio/swen-game-narrative,biblio/swen-phil-lit,biblio/jjb}  

\begin{thebibliography}{15}
\providecommand{\natexlab}[1]{#1}
\providecommand{\url}[1]{\texttt{#1}}
\providecommand{\urlprefix}{}

\bibitem[{Brooks(1986)}]{brooks_robust_1986}
Brooks, R.: A robust layered control system for a mobile robot.
\newblock Robotics and Automation, {IEEE} Journal of 2(1), 14--23 (1986)

\bibitem[{Bryson(2001)}]{Bryson-PHD}
Bryson, J.J.: Intelligence by Design: {P}rinciples of Modularity and
  Coordination for Engineering Complex Adaptive Agents.
\newblock Ph.D. thesis, {MIT}, Department of {EECS}, Cambridge, MA (June 2001), 2001-003

\bibitem[{Champandard and Dunstan(2013)}]{Champandard2013}
Champandard, A.J., Dunstan, P.: The behavior tree starter kit.
\newblock In: Rabin, S. (ed.) Game AI Pro: Collected Wisdom of Game AI
  Professionals, pp. 72--92. Game Ai Pro, A. K. Peters, Ltd. (2013)

\bibitem[{Dias et~al.(2014)Dias, Mascarenhas, and Paiva}]{dias2014fatima}
Dias, J., Mascarenhas, S., Paiva, A.: Fatima modular: Towards an agent
  architecture with a generic appraisal framework.
\newblock In: Emotion Modeling, pp. 44--56. Springer (2014)

\bibitem[{Gaudl(2016)}]{Gaudl-PHD}
Gaudl, S.E.: Building Robust Real-Time Game AI: Simplifying \& Automating
  Integral Process Steps in Multi-Platform Design.
\newblock Ph.D. thesis, Department of Computer Science, University of Bath
  (2016)

\bibitem[{Gaudl et~al.(2013)Gaudl, Davies, and Bryson}]{Gaudl2013}
Gaudl, S.E., Davies, S., Bryson, J.J.: Behaviour Oriented Design for
  real-time-strategy games.
\newblock In: Proceedings of the Foundations of Digital Games. pp. 198--205.
  Society for the Advancement of Science of Digital Games (2013)

\bibitem[{Grow et~al.(2014)Grow, Gaudl, Gomes, Mateas, and
  Wardrip-Fruin}]{grow2014methodology}
Grow, A., Gaudl, S.E., Gomes, P.F., Mateas, M., Wardrip-Fruin, N.: A
  methodology for requirements analysis of ai architecture authoring tools.
\newblock In: Foundations of Digital Games 2014. (2014)

\bibitem[{Isla(2005)}]{isla_gdc_2005}
Isla, D.: {GDC} 2005 proceeding: Handling complexity in the halo 2 {AI}.
\newblock

  (2005)
  

\bibitem[{Keith(2010)}]{keith2010agile}
Keith, C.: Agile Game Development with Scrum.
\newblock Addison-Wesley Signature Series (Cohn), Pearson Education (2010),

\bibitem[{Mateas(2002)}]{Mateas2002a}
Mateas, M.: Interactive Drama, Art, and Artificial Intelligence.
\newblock Technical report cmu-cs-02-206, School of Computer Science, Carnegie
  Mellon University (December 2002)

\bibitem[{Mateas and Stern(2002)}]{Mateas2002}
Mateas, M., Stern, A.: A behavior language for story-based believable agents.
\newblock Intelligent Systems, IEEE 17(4), 39--47 (2002)

\bibitem[{Orkin(2005)}]{orkin2005agent}
Orkin, J.: Agent architecture considerations for real-time planning in games.
\newblock In: Young, M.R., John, L. (eds.) Proceedings of the First Artificial
  Intelligence and Interactive Digital Entertainment Conference. pp. 105--110.
  AAAI Press, Menlo Park, CA (2005)

\bibitem[{Partington and Bryson(2005)}]{Partington05}
Partington, S.J., Bryson, J.J.: The {B}ehavior {O}riented {D}esign of an
  {U}nreal {T}ournament character.
\newblock In: Panayiotopoulos, T., Gratch, J., Aylett, R., Ballin, D., Olivier,
  P., Rist, T. (eds.) The Fifth International Working Conference on Intelligent
  Virtual Agents. pp. 466--477. Springer, Kos, Greece (September 2005)

\bibitem[{Rubin(2012)}]{rubin2012essential}
Rubin, K.: Essential Scrum: A Practical Guide to the Most Popular Agile
  Process.
\newblock Addison-Wesley Signature Series (Cohn), Pearson Education (2012),

\bibitem[{Spierling and Szilas(2009)}]{spierling2009authoring}
Spierling, U., Szilas, N.: Authoring issues beyond tools.
\newblock In: Joint International Conference on Interactive Digital
  Storytelling. pp. 50--61. Springer (2009)

\end{thebibliography}
	%
	%
\end{document}